\documentclass{article}
\usepackage[english]{babel}
\usepackage{amsfonts,amssymb, amsmath}

\newtheorem{prop}{Proposition}
\newtheorem{thr}{Theorem}
\newtheorem{defi}{Definition}

\DeclareMathOperator{\rank}{rank}

\DeclareMathOperator{\dv}{div}

\newcommand{\Lie}{\mathcal{L}}
\newcommand{\dd}{\mathrm{d}}
\newcommand{\pd}{\partial}

\begin{document}

\title{From Jacobi and Lie to tensor invariants: two-term complete systems and Poisson geometry}
\author{A.\,V.~Tsiganov\\
\it\small International Laboratory for Mirror Symmetry\\ 
\it\small National Research University Higher School of Economics, 
Moscow, Russia\\
\it\small Beijing Institute of Mathematical Sciences and Applications, Beijing, China\\
\it\small email: andrey.tsiganov@gmail.com}

\date{}

\maketitle

\begin{abstract}
A smooth involutive distribution $D\subset TM$ is integrable which allows us to construct global tensor invariants of the flows associated to a vector field $X\in D$. The corresponding scalar conserved quantities may be multivalued  functions on $M$. As an example we consider the case $\rank D= 2$ in detail.

\end{abstract}

\section{Introduction}
\setcounter{equation}{0}
In 1918, Emmy Noether published her famous article on invariant variational problems in which she stated and proved the so-called Noether theorem \cite{no-18}.  In modern physics, Noether's theorem is most commonly used to describe the relationship between global continuous symmetries and conserved quantities. Interestingly, Noether herself never used the word 'symmetry' in her paper. Instead, she discusses integrals that are mathematically analogous to the action integrals of Lagrangian physics. Using variational techniques and group theory, she establishes a one-to-one correspondence between the variational symmetries of the integral and a set of identities.

In 1827 Jacobi described similar relationship between a commuting set of symmetries and conserved quantities in the framework of the Pfaff theory, Hamiltonian mechanics and Poisson geometry \cite{jac-27,jac-27a}. The Jacobi results were generalised to  the non-commutative case by Deahna \cite{deah-40}, Clebsch \cite{cl-66}, Frobenius \cite{frob-77}, Lie \cite{lie-77}, Mayer \cite{may-72} among many others, see textbooks \cite{forth-06,car}.  All of these theorems are local in nature, so conserved quantities are also locally defined in the generic case.

The scalar conservative quantities may be local, multivalued, or transcendental functions following  Euler's terminology \cite{eul-int}. The corresponding tensor conservative quantities may be single-valued  (polynomial) tensor fields. For instance, we present  polynomial Poisson bivector with the transcendental Casimir function associated with the Darboux-Halphen differential system.

In \cite{bog96,bog97} Bogoyavlenskij proposed the construction of tensor invariants in the neighbourhood of the Liouville tori of a completely integrable Hamiltonian system. According to \cite{ts25,ts25a,ts25b,ts25c,ts25ar}, some of these invariants are globally defined. Furthermore, global tensor invariants have been found for a number of systems that are not integrable according to Liouville's theorem.

In this note we discuss single-valued tensor conservative quantities for  ho\-mo\-ge\-ne\-ous, scale invariant and weight invariant Hamiltonian vector fields. All of these invariants can be obtained easily within the framework of the well-known $r$-term complete differential systems theory. As an example case $r=2$ is considered in detail.

\subsection{Example: basic tensor invariants in Hamiltonian mechanics}

A vector field $X$ that appears in the Hamilton equations
\begin{equation}\label{hg-eq}
\dot{x}_k=X_k(x_1,\ldots,x_n)\,,\qquad k=1,\ldots,n\,,
\end{equation}
is uniquely determined by the  following equations
\begin{equation}\label{h2-eq}
X=P \dd H\qquad \mbox{or}\qquad \iota_X \omega=-\dd H\,,
\end{equation} 
The basic tensor invariants of equations (\ref{hg-eq},\ref{h2-eq}) are the Hamilton function $H$, Hamiltonian vector field $X$, the Poisson bivector $P$ or the symplectic form $\omega$. 

A smooth tensor field $T(x)$ is called a tensor invariant of equations (\ref{hg-eq}) if
\begin{equation*}
\mathcal L_XT(x) = 0\,.
\end{equation*}
Here $\mathcal L_X$ is the Lie derivative of the field $T(x)$ along the vector field $X$, which determines the rate of change of the tensor field $T(x)$ under the deformation of the phase space defined by the flow of $X$. In local coordinates the Lie derivative of the tensor field $T$ of type $(p, q)$ is equal to
\begin{align*}
({\mathcal {L}}_{X}T)^{i_{1}\ldots i_{p}}_{j_{1}\ldots j_{q}}=&\sum_{k=1}^n X^{k}(\partial _{k}T^{i_{1}\ldots i_{p}}_{j_{1}\ldots j_{q}})\\
-&\sum_{\ell=1}^n (\partial _{\ell}X^{i_{1}})T^{\ell i_{2}\ldots i_{p}}_{j_{1}\ldots j_{q}}-\ldots - \sum_{\ell=1}^n (\partial _{\ell}X^{i_{p}})T^{i_{1}\ldots i_{p-1}\ell}_{j_{1}\ldots j_{q}}\\
+&\sum_{m=1}^n(\partial _{j_{1}}X^{m})T^{i_{1}\ldots i_{p}}_{m j_{2}\ldots j_{q}}+\ldots +\sum_{m=1}^n(\partial _{j_{q}}X^{m})T^{i_{1}\ldots i_{p}}_{j_{1}\ldots j_{q-1}m}
\nonumber
\end{align*}
where $\partial _s= {\partial }/{\partial x^s}$ is the partial derivative on the $x^s$ coordinate. Fortunately, we can prove the invariance of the  basic tensor invariants of Hamiltonian flows (\ref{hg-eq},\ref{h2-eq})
without having to perform any calculations in local coordinates.

The vector field $X$ is a tensor invariant because of the antisymmetry of the Lie bracket $[.,.]$
\[
\mathcal L_X\,X=[X,X]=0\,.
\]
The Hamilton function $H$ is a tensor invariant because Poisson bivector $P$ is a skew-symmetric tensor field
\[
\mathcal L_X\,H=(\dd H,P\dd H)=0\,.
\]
The Poisson bivector $P $ is a tensor invariant 
\begin{equation}\label{inv-gen}
\mathcal L_X\,P=\mathcal L_{P\dd H}\,P=\frac{1}{2}[\![P,P]\!]\, \dd H=0
\end{equation}
by virtue of the Jacobi identity
\begin{equation}\label{jac-eq}
[\![P,P]\!]=0\,,
\end{equation}
which we write using the Schouten brackets $[\![.,.]\!]$.

Obviously, the validity of the Jacobi identity (\ref{jac-eq}) is a sufficient but not necessary condition for the invariance equation (\ref{inv-gen}) to hold. For example, the bivector
\[\tilde{P}=(a_1H+a_2)P\,,\qquad a_{1,2}\in\mathbb R\,,\]
does not satisfy the Jacobi identity for $a_1\neq0$, but nevertheless is a tensor invariant for arbitrary values of the parameters $a_{1,2}$.

The invariance of the symplectic form is a consequence of $\dd \omega=0$ and the Hamilton equations (\ref{h2-eq}), since
\[
\mathcal L_X \omega=\dd(\iota_{ X}\omega) + \iota_{X}(\dd\omega) = \dd(-\dd H) = 0\,.
\]
This leads to the conservation of the $2n$ volume form
\[
\Omega=\wedge^n \omega\,,\qquad \mathcal L_X\Omega=0\,.
\]
As an example, consider the Hamiltonian system of differential equations
\begin{equation}\label{h-eq}
\frac{dq_i}{dt}= \frac{\partial H}{\partial p_i}\,,\qquad
\frac{dp_i}{dt}= -\frac{\partial H}{\partial q_i}\,,\qquad i=1,\ldots,n,
\end{equation}
where $q=\{q_1,\ldots,q_n\}$ are coordinates in configuration space $\mathbb R^n$,  $p=\{p_1,\ldots,p_n\}$ are the corresponding momenta in phase space $T^*\mathbb R^n$, and $H=H(q,p)$ is the Hamilton function.

In local coordinates $x=(q,p)$ vector fields read as
\begin{equation}\label{x-vec}
X=\sum_{i=1}^n \left( \frac{p_i}{m_i} \frac{\partial}{\partial q_i}- \frac{\partial V}{\partial q_i}\dfrac{\partial}{\partial p_i}\right)\,.
\end{equation}
The Poisson bivector $P$ and symplectic form $\omega$ have the form 
\begin{equation}\label{p-can}
P=\sum_{i=1}^n\frac{\partial}{\partial q_i}\wedge\frac{\partial}{\partial p_i}\qquad\mbox{and}\qquad
\omega=\sum_{i=1}^n dp_i\wedge dq_i\,,
\end{equation}
so that
\[
\Omega=\wedge^n \omega=(-1)^{n(n+1)/2}n!\, dq_1\wedge\cdots\wedge dq_n\wedge dp_1\wedge\cdots\wedge dp_n\,.
\]
The corresponding invariant Poisson brackets on the phase space $T^*\mathbb R^n$ are equal to
\[
\{q_i,p_j\}=\delta_{ij}\,,\qquad \{q_i,q_j\}=\{p_i,p_j\}=0\,,\qquad i,j=1,\ldots,n.
\]
Some other tensor invariants of certain Hamiltonian systems were obtained by solving the invariance equation 
$\mathcal L_XT=0$ directly in local coordinates, see \cite{ts25,ts25a,ts25b,ts25c,ts26a} 

\section{Complete systems of differential equations}
\setcounter{equation}{0}

Ever since the late 18-th century mathematicians had realized that the integration of
a single linear homogeneous partial differential equation
\begin{equation}\label{pde-jac}
A(f)=0\,,\qquad A(f)=\sum_{j=1}^n \xi_j(x) \frac{\partial f}{\partial x_j}
\end{equation}
was equivalent to the integration of a system of first order ordinary differential equations
\begin{equation}\label{ode-jac}
\frac{\dd x_1}{\xi_1}=\cdots=\frac{\dd x_n}{\xi_n}\,.
\end{equation}
In 1827   Jacobi gave a treatment of this equivalence \cite{jac-27,jac-27a}. In 1872  Mayer used equivalence between (\ref{pde-jac}) and (\ref{ode-jac}) to establish a reciprocal connection between systems
of linear homogeneous partial differential equations and Pfaffian systems \cite{pfaff-15}
\[
\omega_i=0\,,\qquad \omega_i=\sum_{i,j=1}^n a_{ij}(x)\dd x_j\,,
\]
 which "in
particular cases ... has already been observed and utilized many times" \cite{may-72}. The Mayer results were generalized by Lie \cite{lie-77} and Frobenius \cite{frob-77}.

We briefly discuss relations between (\ref{pde-jac}) and (\ref{ode-jac}) following to Lie's textbook \cite{lie-book}.
Suppose that the functions $\xi_{1}, \dots, \xi_{n}$ are different from zero at a point $x^{0}=(x_{1}^{0}, \dots, x_{n}^{0})$  and  behave regularly in the neighborhood.  Under these assumptions, we can determine $x_{1}, \dots, x_{n-1}$ as
analytic functions of $x_{n}$ in such a way that, by substituting these functions into the equations
\begin{equation}\label{sim-eq}
  \frac{dx_{1}}{dx_{n}} = \frac{\xi_{1}}{\xi_{n}}, \ \dots,\ 
  \frac{dx_{n-1}}{dx_{n}} = \frac{\xi_{n-1}}{\xi_{n}}
\end{equation}
is identically satisfied and that in addition $x_{1}, \dots, x_{n-1}$ for
$x_{n} = x_{n}^{0}$ take certain prescribed {initial values}
 $x'_{1}, \dots, x'_{n-1}$ which are interpreted as the integration constants.

\begin{defi}
The equations representing  $x_{1}, \dots, x_{n-1}$
as functions of $x_{n}$
\[x_{k}=\phi_k(x_n,x'_1,\ldots,x'_{n-1}),\qquad k=1,\ldots,n-1 
\]
are called the complete integral equations of (\ref{sim-eq}). The ordinary differential equations (\ref{ode-jac}) are completely integrable when $n-1$ functionally independent functions  $\phi_k (x_n,x'_1,\ldots,x'_{n-1})$  are known.
\end{defi}
  Of course, it is only local definition in the neighborhood of a point satisfying the specified conditions.

In  Lie's textbook \cite{lie-book}, such local solutions $\phi_k (x_n,x'_1,\ldots,x'_{n-1})$ were represented as follows
\[
  x_{k} = x'_{k} + (x_{n}^{0} - x_{n})\,
  \mathfrak{P}_{k}\big(x'_{1} - x_{1}^{0}, \dots, x'_{n-1} - x_{n-1}^{0},
              x_{n}^{0} - x_{n}\big)
  \qquad k = 1,\ldots,n-1,
\]
where the $\mathfrak{P}_{k}$ denote ordinary power series in their arguments. Inverting the relationship between $x_1, \ldots, x_n$ and $x_1',\ldots, x_n'$ we obtain 
\[
  x'_{k} = x_{k} + (x_{n} - x_{n}^{0})\,
  \mathfrak{P}_{k}\big(x_{1} - x_{1}^{0}, \dots, x_{n} - x_{n}^{0}\big)
  = f_{k}(x_{1}, \dots, x_{n})
  \qquad k = 1,\ldots, n-1.
\]
\begin{defi}
The functions $f_{k}$ are the so-called integral functions 
of the ordinary differential equations (\ref{ode-jac}), since the differentials of these functions
\[
  df_{k} = \sum_{i=1}^{n} \frac{\partial f_{k}}{\partial x_{i}}\, dx_{i}=0
\]
all vanish identically by virtue of (\ref{sim-eq}). 
\end{defi}
Of course, we  suppose that $f_k = x'_{k}$ can be  differentiated with respect to $x_n$ and, therefore, each function on $z =f_k(x_1,\ldots,x_n)$ is then a local solution to partial differential equation (\ref{pde-jac}) which is also called complete equation in this case.

In his new method \cite{jac}, Jacobi showed that when the $r<n$ equations
\begin{equation}\label{pde-jac-sys}
A_i(f)=0\,,\qquad i=1,\ldots,r\,,\qquad  A_i(f)=\sum_{j=1}^n \xi_{ij}(x) \frac{\partial f}{\partial x_j}
\end{equation}
are linearly independent and satisfy the integrability condition
\begin{equation}\label{jac-int}
A_i\bigl(A_j(f)\bigr)-A_j\bigl(A_i(f)\bigr)=0\,,\qquad \mbox{for all}\quad i\neq j\,,
\end{equation} 
then (\ref{pde-jac-sys} ) is a complete system with $n - r$ independent solutions $f_1,\ldots, f_{n-r}$.
For systems satisfying (\ref{jac-int}) Jacobi sketched a method for finding the solutions
via integration of systems of ordinary differential equations using Poisson brackets, see discussion in \cite{deah-40,cl-66,frob-77} and textbooks \cite{forth-06,lie-book}.

In \cite{cl-66} Clebsch proposed the following generalization of the Jacobi results.
\begin{thr}
A system of $r < n$ linearly independent first order partial
differential equations (\ref{pde-jac-sys}) in $n$ variables is complete in the sense that there
exist $n- r$ functionally independent solutions $f_1,\ldots,f_{n-r}$ to the system if and only if
\begin{equation}\label{cl-int}
A_i\bigl(A_j(f)\bigr)-A_j\bigl(A_i(f)\bigr)=\sum_{k=1}^r \chi_{ijk}(x) A_k(f)\,.
\end{equation}
Here the functions $ \chi_{ijk}(x)$ satisfy  the Jacobi's identity.
\end{thr}

To prove Clebsch's theorem  Mayer  solved an original $r$-term differential system
\[
  A_{1}(f) = 0, \ldots, A_{r}(f) = 0
\]
with respect to $r$ of the differential quotients, then the resulting equations
\begin{equation}\label{lie-qterm}
  B_{k}(f) = \frac{\pd f}{\pd x_{n-r+k}}
  + \sum_{i=1}^{n-r} \eta_{ki}\, \frac{\pd f}{\pd x_{i}} = 0\,,\qquad  k = 1,\ldots,r\,,
\end{equation}
stand pairwise in the relationships
\begin{equation*}
  B_{i}\big(B_{k}(f)\big) -B_{k}\big(B_{i}(f)\big) = 0\,,\qquad i, k = 1,\ldots, r\,.
\end{equation*}
According to Lie  the system (\ref{lie-qterm}) possesses $n - r$ independent solutions,
for the determination of which it suffices to integrate $r$ individual linear
partial differential equations one after the other \cite{lie-book}.

After that Lie formulate his main theorem of his theory of complete differential systems.
\begin{thr}
Every $r$-term complete system:
\[
  \frac{\pd f}{\pd x_{n-r+k}}
  + \sum_{i=1}^{n-r} \eta_{ki}(x_{1}, \dots, x_{n})\, \frac{\pd f}{\pd x_{i}}
  = 0\,,   \qquad k = 1,\ldots,r
\]
whose coefficients $\eta_{ki}$ behave regularly in the neighborhood of
$x_{1} = x_{1}^{0}$, \dots, $x_{n} = x_{n}^{0}$ possesses $n - r$ independent
solutions $x_{1}^{(r)}, \dots, x_{n-r}^{(r)}$ which behave regularly in a
certain neighborhood of $x_{1} = x_{1}^{0}$, \dots, $x_{n} = x_{n}^{0}$ and
which in addition reduce to $x_{1}, \dots, x_{n-r}$ respectively after the
substitution $x_{n-r+1} = x_{n-r+1}^{0}, \dots, x_{n} = x_{n}^{0}$.
\end{thr}
Detailed proofs of the Clebsch and Lie theorems can be found in Lie's textbook \cite{lie-book}.

\subsection{Example: construction of symmetries}
A vector field $X$ on a manifold $M$ has a symmetry if there exists a transformation $\varphi: M\to M$ such that applying $\varphi$ to $X$ gives back the same field (or a related one) \cite{mm-21}. If $M$ has a metric $g$, a metric symmetry of the vector field that preserves the metric is generated by a Killing vector field $K$
\[
\mathcal L_K g=0\,.
\]
If additionally $[K,X]=0$, then $K$ is a symmetry of $X$ \cite{mm-21}. Example: in a spherically symmetric spacetime, the angular momentum vector fields are Killing fields, and a radial vector field is invariant under rotations.

A vector field $K$ is conformal if \[\mathcal L_K g=2\sigma g\,,\] where $\sigma$ is a scalar function \cite{mm-21}.
A vector field $X$ can be conformally invariant if its Lie derivative along $K$ is proportional to itself
\[\mathcal L_KX=\chi(x) X\,.\]
Example: a scale invariance and angle preservation matter  from 2D CFT and string theory to cosmology and the conformal bootstrap.

For a vector field $X$ generating a flow  $\dot{x}=X$ a symmetry is a transformation of the variables that maps solutions to solutions \cite{car,lie-book}. The infinitesimal generator of the Lie symmetry  $K$ satisfies
\[ [K,X]=\chi(x) X \]
for some function  $\chi(x)$. The most famous examples are the isotropic harmonic oscillator and hydrogen atom \cite{kl-74}.

Obviously, isometries (metric symmetries), Lie symmetries (dynamical symmetries)  and conformal symmetries are partial cases of the vector fields satisfying Clebsch's conditions (\ref{cl-int}) that allows us to relate these symmetries with integrals $f_k$, i.e. solutions of the equations (\ref{pde-jac-sys}).

An $r$-term complete system in $n$ independent variables $A_i(f)=0$ with 
\[
[A_i,A_j]=\sum_{k=1}^r \chi_{ijk}(x) A_k
\]
always possesses $n - r$ independent solutions if and only if functions $\chi_{ijk}(x)$ satisfy the Jacobi identity.  
In \cite{lie-book} Lie also proved the inverse statement.
\begin{thr}
If $r$ independent linear partial differential equations of first order in
$n$ independent variables $A_i(f)=0$ have exactly $n - r$ independent solutions in common, then they form an
$r$-term complete system.  
\end{thr}
It means that functions $\chi_{ijk}(x)$ in the corresponding Clebsch's relations (\ref{cl-int}) satisfy the Jacobi identity. 

Let us consider 3D vector field associated with Euler's top
\[
A_1=a_1x_2x_3\frac{\pd}{\pd x_1}+a_2x_3x_1\frac{\pd}{\pd x_2}+a_3x_1x_2\frac{\pd}{\pd x_3}\,.
\] 
By adding generator of dilations 
\[
A_2=x_1\frac{\pd}{\pd x_1}+x_2\frac{\pd}{\pd x_2}+x_3\frac{\pd}{\pd x_3}\,,
\]
we obtain 2-term complete system with one independent integral 
\[
 A_1(f_{12})=0,\quad A_2(f_{12})=0\,,\qquad\mbox{where}\qquad f_{12}=\frac{a_2x_3^2-a_3x_2^2}{a_3x_1^2-a_1x_3^2}\,. 
\]
By adding the following vector field
\[
A_3=\frac{1}{x_1}\frac{\pd}{\pd x_1}+\frac{1}{x_2}\frac{\pd}{\pd x_2}+\frac{1}{x_3}\frac{\pd}{\pd x_3}
\]
we obtain two complete  differential systems 
\[
A_1(f_{13})=0,\quad A_3(f_{13})=0\,,\qquad \mbox{where}\qquad
f_{13}=a_1(x_2^2 - x_3^2) +a_2(x_3^2 - x_1^2) + a_3(x_1^2 - x_2^2)\,,
\]
and
\[
A_2(f_{23})=0,\quad A_3(f_{23})=0\,,\qquad \mbox{where}\qquad
f_{23}=\frac{x_2^2 - x_3^2}{x_1^2 - x_2^2}\,.
\]
The corresponding Lie's brackets have the following form 
\[ 
[A_1,A_2]=-A_1\,,\qquad
[A_1,A_3]=-\left(\frac{1}{x_1^2}+\frac{1}{x_2^2}+\frac{1}{x_3^2} \right)A_1\,,\qquad [A_2,A_3]=-2A_3\,.
\]
It is easy to verify that Jacobi's condition holds 
\[
[A_1,[A_2,A_3]]+[A_2,[A_3,A_1]]+[A_3,[A_1,A_2]]=0\,.
\]
As it is impossible to classify systems of complete equations $A_i(f)=0$ when Clebsch's conditions
\[
[A_i,A_j]=\sum_{k=1}^r \chi_{ijk}(x) A_k\,,\qquad \forall i,j=1,\qquad r\,,
\]
involve functions $\chi_{ijk}(x)$, even when the homogeneity condition is imposed, Lie limits his discussion in the following sections of his textbook \cite{lie-book} to the case constant real coefficients $\chi_{ijk}$ which satisfy  the Jacobi's conditions.  As a result instead of generic Lie theorem we have a partial theorem about solvable Lie's algebras \cite{koz-24}.

\subsection{Example: construction of new complete systems}
Let us consider 2-term complete differential system defined by generator of dilatations 
\[
A_1=x_1\frac{\pd}{\pd x_1}+x_2\frac{\pd}{\pd x_2}+x_3\frac{\pd}{\pd x_3}
\]
and vector field
\[
A_2=a_1x_1^\kappa \frac{\pd}{\pd x_1}+a_2x_2^\kappa \frac{\pd}{\pd x_2}+a_3x_3^\kappa \frac{\pd}{\pd x_3}\,,
\]
so that $[A_1,A_2]=(\kappa-1)A_2$\,. Functions
\[
d_{i}=a_jx_k^{1-\kappa}-a_kx_j^{1-\kappa}\,,\qquad i\neq j\neq k\neq i
\]
are solutions of equation $A_2(d_i)=0$ so that $A_1(d_m/d_n)=0$. So, according to Clebsch's theorem two term complete system $A_1(f)=0$, $A_2(f)=0$ has a one independent integral.

Let us add to this system third partial differential equation 
\[A_3(f)=0\,,\qquad
A_3=\phi_1(x)\frac{\pd}{\pd x_1}+\phi_2(x)\frac{\pd}{\pd x_2}+\phi_3(x)\frac{\pd}{\pd x_3}
\]
which contains three unknown functions $\phi_k(x)$. If we suppose that Lie's brackets form a simple real algebra
\[
[A_1,A_2]=(\kappa-1)A_2\,,\qquad
[A_1,A_3]=-(\kappa-1)A_3\,,\qquad
[A_2,A_3]=2(\kappa-1)A_1\,,
\]
which is isomorphic to $\mathfrak{sl}(2)$, then functions $\phi_i$ have to be
\[
\phi_i(x)=\frac{x_i^{2 - \kappa}}{a_i} + x_i^\kappa \,d_i^2\, F_i\left(\frac{d_m}{d_n}\right),\qquad i=1,2,3,
\]
where $F_i$ are arbitrary functions on a common solution $d_m/d_n$ of $A_1(f)=0$ and $A_2(f)=0$. 
 For a special choice of the functions $F_k$ and at $\kappa = 0$, we obtain the Darboux–Halphen differential system, also known as the Ramanujan vector field, see details in \cite{ts26d}.

Since Lie's determinant \cite{lie-1}
\[
\Delta=\left\|
  \begin{array}{ccc}
    x_1 & x_2 & x_3 \\
    a_1x_1^\kappa & a_2x_2^\kappa & a_3x_3^\kappa \\
    \phi_1 & \phi_2 & \phi_3 \\
  \end{array}
\right\|
\]
defines the Jacobi multiplier $M=\Delta^{-1}$ for all the vector fields, i.e.
\[
\dv (MA_1)=0\,,\qquad  \dv (MA_2)=0\,,\qquad \dv (MA_3)=0\,,
\]
we have complete integrable differential systems, according to the Euler-Jacobi theorem.

In a similar way we can obtain complete integrable differential systems associated with other well-known 3D real Lie algebras \cite{ts26rcd}.

\section{Homogeneous vector fields}
\setcounter{equation}{0}
A function  $f(x_1,\dots,x_n)$ is homogeneous of degree $\kappa$ if 
\[f(\lambda x_1,\ldots,\lambda x_n)=\lambda^\kappa f(x_1,\ldots,x_n)\qquad \forall\,\lambda>0\,.\] 
The Euler's theorem states
\[
E(f)\equiv\sum_{i=1}^n x_i \frac{\partial}{\partial x_i}f(x_1,\ldots,x_n)=\kappa f(x_1,\ldots,x_n)\,.
\]
The corresponding Euler vector field 
\begin{equation}\label{scale-euler}
E=\sum_{i=1}^n x_i\frac{\partial}{\partial {x_i}}
\end{equation} 
is the infinitesimal generator of scale transformations or dilations.

Let us consider a smooth vector field
\[
X=\sum_{i=1}^n X_i(x_1,\ldots,x_n)\frac{\partial}{\partial x_i} \,,
\]
whose components $X_i(x_1,\ldots,x_n)$ are homogeneous functions of the variables $x_1,\ldots,x_n$ of homogeneity degree $\kappa$
\[
X_i(\lambda x_1,\ldots,\lambda x_n)=\lambda^\kappa X_i(x_1,\ldots, x_n)\,,\qquad i=1,\ldots,n\,.
\]
The Lie bracket between  vector fields $E$ and $X$ is equal to
\[
[E,X]=(\kappa-1)X\,.
\]

\begin{prop}
The following Poisson bivector
\begin{equation}\label{my-p}
P_h =X\wedge E
\end{equation}
is a tensor invariant of the phase flow defined by  $X$.
\end{prop}
We verify this Proposition by computing Lie derivative of $P_h$ along  $X$
\[\mathcal L_X\,P_h =[X,X]\wedge E+X\wedge [X,E]=0\,,\]
and  to check the validity of the  Jacobi identity
\[
[\![P_h,P_h]\!]=-2X\wedge [E,X]\wedge E=0\,.
\]
So, the phase flow of a homogeneous vector field $X$ preserves the Poisson bivector $P_h$
associated with the scale transformations (\ref{scale-euler}) \cite{ts26a}.

\subsection{Casimir functions and Hamiltonian  vector fields} 
A function $f$ is called a Casimir function of the Poisson bivector $P$ if $Pdf=0$. The Poisson bivector $P_h$ (\ref{my-p}) has at least $n-2$ Casimir functions, since the vector fields $E$ and $X$ satisfy the Clebsch integrability condition (\ref{cl-int}) 
\[[E, X] = (\kappa - 1)X\] 
for any $\kappa$.
\begin{prop}
According to Clebsch's theorem a pair of partial differential equations 
\[E(f) = 0\qquad \mbox{and}\qquad X(f) = 0\]
has $n - 2$ functionally independent solutions $f_1,\ldots,f_{n-2}$ which are the Casimir functions of the Poisson bivector $P_h$ (\ref{my-p}).  
\end{prop}
The following direct computation verifies this Proposition
\[
P_h\dd f_k=-X(f_k)\,E+E(f_k)\,X=0\,.
\]

Suppose now that the function $F(x_1,\ldots,x_n)$ is a homogeneous first integral of a vector field $X$ of homogeneity degree $\varkappa\neq 0$, that is,
\[
X(F)=\sum_{i=1}^n X_i\frac{\partial F}{\partial x_i}=0\qquad\mbox{and}\qquad
E(F)=\sum_{i=1}^n x_i\frac{\partial F}{\partial x_i}=\varkappa F\,.
\]
In this case
\[
P_hdF=-X(F)\,E+E(F)\,X=\varkappa F\, X\,,
\]
that allows us to represent a homogeneous vector field $X$ in Hamiltonian form  (\ref{h2-eq}) at $\varkappa\neq 0$
\[
X=P_h\dd H\,,\qquad H=\varkappa^{-1}\ln |F|\,.
\]
\subsection{Example: 3D homogeneous  Hamiltonian vector fields}
Let us consider a vector field with second-order polynomial entries
\begin{equation}\label{g-lv}
X_i=\sum_{j,k=1}^n a_{ijk}x_jx_k\,,\qquad i=1,\ldots,n\,,\qquad a_{ijk}\in\mathbb R\,. 
\end{equation}
The corresponding Poisson bivector $P_h=X\wedge E$ (\ref{my-p}) defines cubic Poisson brackets
\[
\{x_i,x_j\}=(x_iX_j-x_jX_i)\,.
\]
which are invariant to the flow of $X$.

If we assume that flow of $X$ preserves linear Hamiltonian
\[
H_{1}=x_1+\cdots+x_n
\]
we obtain linear algebraic equations on coefficients $a_{ijk}$ in (\ref{g-lv}). Solving these equations we obtain Hamiltonian vector field
\[
X_{1}=-P_hd \ln H_{1}\,.
\]
The partial case of such systems is an integrable generalized Lotka–Volterra systems.

If we assume that flow of $X$ preserves quadratic Hamiltonian
\[
H_{2}=x_1^2+\cdots+x_n^2
\]
we obtain another set of  linear equations on coefficients $a_{ijk}$ (\ref{g-lv}). Solving these equations we obtain another Hamiltonian vector field
\[
X_{2}=-\frac{1}{2}P_hd \ln H_{2}\,.
\]
For instance, at $n=3$ we have the following Hamiltonian vector field
\[
X_{2}=
\left(
  \begin{array}{c}
    (ax_2 + bx_3)x_1 + x_2^2c + d + f) x_2x_3 + ex_3^2 \\
    -ax_1^2 - (cx_2 + dx_3)x_1 + gx_2x_3 + hx_3^2 \\
    -bx_1^2 - (ex_3 + fx_2)x_1 - gx_2^2 - hx_2x_3 \\
  \end{array}
\right)
\]
depending on eight parameters $a,\ldots,h$. 

At $a=h$, $b=-g$ and $c=-e$ the Jacobi multiplier equals to unit $M=1$, div$X_2=0$, and we have completely integrable Hamiltonian system according to the Euler-Jacobi theorem. 

\subsection{Example: transcendental Casimir function}
Let us consider the Darboux-Halphen system 
\begin{equation}\label{dh-sys}
u'_2+u'_3=u_2u_3\,,\qquad
u'_3+u'_1=u_1u_3\,,\qquad
u'_1+u'_2=u_1u_2,,\qquad u'=\frac{\dd u}{\dd x}
\end{equation}
In \cite{bri} Brioschi  used projective coordinates to reduce these equations to a system of ordinary differential equations of second order
\[
\frac{\dd^2 y}{\dd x^2}=0\qquad \mbox{and}\qquad \dfrac{\dd^2 w}{\dd \xi^2}+\mathrm{P}\,\frac{\dd w}{\dd \xi}+\mathrm{R}\,w=0\,, 
\]
see equations (5) in \cite{bri}. The pairs of their independent solutions are expressed in terms of hypergeometric functions 
\[
 y_{0}=F\left(\alpha,\beta,\gamma,x\right)\,,\quad y_{1}=F\left(\alpha,\beta,\alpha+\beta-\gamma+1,1-x\right)\,,
\]
and
\[ w_{0}=F\left(\frac12,\frac12,1,\xi\right)\,,\quad w_{1}=F\left(\frac12,\frac12,1,1-\xi\right)\,.\]
In terms of complete and complementary complete elliptic integrals of the first kind 
\[
\mathsf{K}(k)=
\int_0^1\frac{1}{ \sqrt{1-\alpha} \sqrt{1-k^2\alpha} } \qquad\mbox{and}\qquad
\mathsf{K}'(k) = \int_0^1
       \frac{1}{\sqrt{1-\alpha}\sqrt{1 +(k^2-1)\alpha}}.
\]
the Brioshi solutions are
\begin{align*}
&y_0=\frac{\mathsf{K}(\sqrt{\xi})}{\sqrt{u_2-u_3}}\,,\qquad
w_0=\sqrt{\xi(\xi-1)}\mathsf{K}(\sqrt{\xi})\,,\qquad 
\qquad
\xi=\frac{u_1-u_3}{u_2-u_3}\,,\\
&y_1=\frac{\mathsf{K}'(\sqrt{\xi})}{\sqrt{u_2-u_3}}\,,
\qquad 
w_1=\sqrt{\xi(\xi-1)}\mathsf{K}'(\sqrt{\xi})\,,
\end{align*}
see original definitions of elliptic integrals in \cite{bri} and modern expressions in \cite{ts26d}.

From the first equation $\dd^2 y/\dd x^2=0$ it follows that $X(F_0)=0$ and $X(F_1)=0$, where $X$ is  the homogeneous vector field defined by equations (\ref{dh-sys}) and
\[
F_0=\frac{\dd}{\dd x}\, y_0\,,\qquad F_1=\frac{\dd}{\dd x} \,y_1\,,\qquad\mbox{so that}\qquad  E(F_{0,1})=\frac{1}{2}\,F_{0,1}\,.
\]
In full accordance with Clebsch's theorem the two-term complete system of partial differential equations 
\[X(f)=0 \qquad\mbox{and}\qquad E(f)=0\] 
has  only one independent solution
\begin{equation}\label{caz-dh}
f=\frac{F_0}{F_1}=\frac{(u_2 - u_3)\mathsf{E}\left(\frac{\sqrt{u_1 - u_3}}{\sqrt{u_2 - u_3}}\right)-u_2\mathsf{K}\left(\frac{\sqrt{u_1 - u_3}}{\sqrt{u_2 - u_3}}\right)}
{
(u_2 - u_3)\mathsf{E}'\left(\frac{\sqrt{u_1 - u_3}}{\sqrt{u_2 - u_3}}\right)
+u_3\mathsf{K}'\left(\frac{\sqrt{u_1 - u_3}}{\sqrt{u_2 - u_3}}\right)}
\end{equation}
where  $\mathsf{E}(z)$ and $\mathsf{E}'(z)$ are complete and complementary complete elliptic integrals of the second kind.

The direct computation verifies that the transcendent function $f$ (\ref{caz-dh}) is the Casimir function of the polynomial Poisson bivector $P_h$ (\ref{my-p})
\[
P_h\dd f=0\,,
\]  
and that Darboux-Halphen vector field $X$ is Hamiltonian 
\[
X=P_h\dd H\,,\qquad H=2\ln F_0\qquad\mbox{or}\qquad H=2\ln F_1\,.
\]

\section{Scale-invariant vector fields}
\setcounter{equation}{0}

The singularity analysis of differential systems proceeds in  a few different steps in order to express all the possible local series expansions around the movable singularities of the solutions. The first step usually consists of finding all possible weight-homogeneous decompositions of the vector field such that the dominant part of the vector field is
a scale-invariant system. The Kowalevski exponents characterize the local behavior of solutions around their scale-invariant solutions \cite{goriely}.  

The class of homogeneous functions can be generalized to the class of weight-homogeneous
functions. A function $f(x_1,\dots,x_n)$ is weight-homogeneous  with
{weights} $w=(w_1,\dots,w_n)$ and {weight degree} $\kappa$ if
\begin{equation}\label{eq:wh}
f(\lambda^{w_1}x_1,\dots,\lambda^{w_n}x_n)=\lambda^{\kappa}f(x_1,\dots,x_n)
\qquad \forall\,\lambda>0 .
\end{equation}
The generalization of Euler's theorem on homogeneous functions is immediate  from
differentiating \eqref{eq:wh} in $\lambda$ at $\lambda=1$. Indeed, let 
\begin{equation}\label{w-eul}
E_w=\sum_{i=1}^{n}w_i x_i\,\dfrac{\partial}{\partial x_i}
\end{equation}
 be the weighted Euler vector
field. Then $f(x_1,\dots,x_n) $ is weight-homogeneous of weight degree $\kappa$ if and only if
\[
E_w(f)=\kappa f(x_1,\dots,x_n) .
\]
A vector field $X$ is weight-homogeneous if 
\begin{equation*}
X_i(\lambda^{w_1}x_1,\dots,\lambda^{w_n}x_n)=\lambda^{\kappa_i}X_i(x_1,\dots,x_n)
\qquad i=1,\ldots,n
\end{equation*}
i.e. when its  Lie brackets with the weighted Euler vector field $E_w$ is equal to
\[
[E_w,X]=WX\,,\qquad W=\mbox{diag}(\kappa_1-w_1,\ldots,\kappa_n-w_n)\,.
\]
The scale-invariant polynomial vector fields with
\begin{equation}\label{scale-inv}
\kappa_i=w_i+c\,,\qquad\Rightarrow\qquad W=cI\,, 
\end{equation}
form an important subclass of weight-homogeneous vector fields. This terminology comes from the invariance of the corresponding system of differential equation under
the scaling transformation
\[t\mapsto\lambda^{-c}t,\qquad x_i\mapsto\lambda^{w_i}x_i\,,\]
where we can set $c=1$ without loss of generality. 
If the corresponding polynomial system
\begin{equation}\label{kow-eq}
\alpha_iw_i = X_i(\alpha_1,\ldots,\alpha_n)\,,\qquad i= 1,\ldots,n
\end{equation}
has at least one non-trivial solution (dominant balance), then there exists a non-trivial particular solution of the form 
\[x_i=\alpha_i t^{w_i}\,,\qquad i=1,\ldots,n\,.\]
In fact, the Kowalevski method consists of studying the local behaviour of solutions around such scale-invariant solutions, see details in \cite{goriely,adam,llibre}.

Not all scale-invariant differential systems have scale-invariant solutions, i.e. solutions of equation (\ref{kow-eq}).
However, for our purposes, there is no need to solve (\ref{kow-eq}) and discuss integrability, since the definition (\ref{scale-inv}) is sufficient for us.
\begin{prop}
The phase flow of a scale-invariant vector field $X$ (\ref{scale-inv}) preserves the Poisson bivector 
\begin{equation}\label{p-w}
P_w=X\wedge E_w\,.
\end{equation}
\end{prop}
Using definition of the scale-invariant vector fields $[E_w,X]=c X$ we have 
\[\mathcal L_X\,P_w=[X,X]\wedge E_w+X\wedge [X,E_w]=0\,,\]
and
\[
[\![P_w,P_w]\!]=-2X\wedge [E_w,X]\wedge E_w=0\,.
\]
So, all the necessary conditions hold.

\subsection{Example: 3D scale-invariant vector field}
Substituting the weighted Euler vector field $E_w$ (\ref{w-eul}) and vector field $X$ with $N$-th order  inhomogeneous polynomial entries  into the Lie brackets
\[
[E_w,X]=X\,,\qquad E_w=\sum_{i=1}^n w_ix_i\,\frac{\partial}{\partial x_i}\,,\qquad 
X=\sum_{k=1}^n X_k\,\frac{\partial}{\partial x_k}
\]
we obtain quadratic-linear equations for weights $w_i$ and coefficients of polynomials $X_k$.

For the fixed values $w_i$ the corresponding linear equations are easily solved. As an example, at $n=3$ and $N=3$ the following choice of  weights
$w_1=1$, $w_2=2$ and $w_3=3$ yields the  scale-invariant vector field
\[
X_w=\left(
      \begin{array}{c}
        a x_1^2 + b x_2 \\
        c x_1^3 + d x_1x_2 + e x_3 \\
        f x_1^2x_2 + g x_1x_3 + h x_2^2 \\
      \end{array}
    \right)
\]
depending on eight parameters $a,\ldots,h$. The flow of $X_w$ preserves Poisson bivector $P_w$ (\ref{p-w}) which can be used to construct Hamiltonian vector fields at the special values of parameters $a,\ldots,h$.

\subsection{Example: Euler - Poisson equations}
The Euler-Poisson equations describing the motion of a rigid body around a fixed point in a homogeneous
gravity field have the form 
\[
\dot{M}=M\times\frac{\pd H}{\pd M}+\gamma\times\frac{\pd H}{\pd \gamma}\,,\qquad
\dot{\gamma}=\gamma\times \frac{\pd H}{\pd M}
\]
where $M=(M_1,M_2,M_3)$, $r = (r_1, r_2, r_3)$, $\gamma=(\gamma_1,\gamma_2,\gamma_3)$ are, respectively, the components of the momentum vector, the components of the radius vector of the center of mass, and the components of the vertical unit vector in the frame of the principal axes. The Hamiltonian $H$ is equal to the full energy of body
\[
H=\frac12 (AM,M)-\mu (r,\gamma)
\]
where $\mu=mg$ is the weight of the body and $A^{-1}$ is the tensor of inertia. This system admits, for all values of the six parameters $r=(r_1,r_2,r_3)$ and $A=diag(a_1,a_2,a_3)$ three time-independent first integrals \cite{goriely}.

The corresponding vector field $X$ is scale-invariant, a fact that Kowalevski exploited in her analysis of the local behaviour of solutions to obtain her seminal results. Using the weighted Euler vector field $E_w$, we can construct the Poisson bivector $P_w = E_w \wedge X$ (\ref{p-w}) with entries of an inhomogeneous cubic polynomial that are invariant under the Euler-Poisson equations.

In \cite{ts25} this independent on integrability bivector was found among the other tensor invariants of Lagrange and Kowalevski tops. The other invariants found in \cite{ts25} are related to complete  integrability  of the Euler-Poisson equations.

\section{Weight-homogeneous decompositions}
\setcounter{equation}{0}
Most vector fields are neither homogeneous nor weight-homogeneous, but many nonlinear
vector fields  can be decomposed into weight-homogeneous components. In general, this weight-homogeneous decomposition is not unique, see geometric interpretation of the decomposition into weight\--\-ho\-mo\-ge\-ne\-ous components  in terms of Newton's polyhedra in the space of exponents in textbook \cite{goriely}.

In classical mechanics decomposing the Hamiltonian 
\[H=T+V\] yields a natural  decomposition of the vector field $X$ (\ref{x-vec})
\[X=P\dd H=P\dd T+P\dd V.\] 
Let us consider the Hamiltonian system of differential equations (\ref{h-eq})
when Hamilton function 
\begin{equation}\label{ham-1}
H=T+V=\frac{1}{2m_1} p_1^2+\cdots+\frac{1}{2m_n} p_n^2+V(q)\,,\qquad m_i>0\,,
\end{equation}
is the sum of the kinetic energy $T$ and a smooth homogeneous function $V(q)$ of homogeneity degree $k$. In this case
the Lagrange identity holds
\begin{equation}\label{eq-lag}
\frac{d^2 J}{dt^2}= 4T - 2kV=4H-2(k+2)V,\qquad J=\sum_{i=1}^n m_iq_i^2\,.
\end{equation}
In  \cite{jac} Jacobi used this identity when studying the stability of the motion of systems 
$n$ interacting particles with masses $m_1,\ldots,m_n$ arising in celestial mechanics, see details in also \cite{alb}.
Various generalisations and applications of the Lagrange identity are discussed in \cite{koz25}.

The necessary conditions for the integrability of systems of equations (\ref{h-eq}) in the class of meromorphic functions are discussed in  \cite{goriely, adam,llibre}.  Throughout this text, we consider Hamiltonian systems in general position, i.e. potentials subject to no conditions other than the homogeneity condition
\[
V(\lambda\,q_1,\ldots,\lambda\,q_n) =\lambda^k\,V(q_1,\ldots,q_n)\,.
\]
According to Euler's theorem for differentiable homogeneous functions, this condition is 
\begin{equation}\label{euler}
Z=0\,,\qquad\mbox{where}\qquad Z=\sum_{i=1}^n q_i\frac{\partial}{\partial q_i}\,V(q_1,\ldots,q_n)-kV(q_1,\ldots,q_n)\,.
\end{equation}
The Hamiltonian (\ref{ham-1}) with a homogeneous potential $V$ of degree $k$  is weight-homogeneous with
\[
\deg q_i=1,\qquad \deg p_i=\frac{k}{2},\qquad \deg H=k
\]
according to Lagrange identity $Z=0$ (\ref{euler}). It allows us to define a weighted dilation generator 
\begin{equation}\label{y-1}
Y=\sum_{i=1}^n q_i\frac{\partial}{\partial q_i}+\frac{k}{2}
\sum_{i=1}^n p_i\frac{\partial}{\partial p_i}
\,,
\end{equation}
associated to scaling 
\[
q\to \lambda q,\qquad p\to \lambda^{k/2}p
\]
under which Hamilton function $H$ is homogeneous
\[
\mathcal L_Y H\equiv Y(H)=k H+Z=kH\,,\qquad \mbox{at}\quad Z=0\,.
\]
At the special value $k=-1$  generator $Y$  is the familiar Kepler scaling $q\to \lambda q$ and $p\to \lambda^{-1/2}p$, whereas at $k=\pm 2$ we have uniform scaling $(q,p)\to \lambda (q,p)$ and a canonical transformation, respectively.

The Liouville form \[\alpha=\sum p_i\,\dd q_i\] has weight $\deg p+\deg q=\tfrac{k}{2}+1$, whence
\[
\Lie_Y\omega=\Bigl(\frac{k}{2}+1\Bigr)\omega,\qquad
\Lie_Y P=-\Bigl(\frac{k}{2}+1\Bigr)P,
\]
and, therefore,
\begin{equation}\label{y-eq}
\mathcal L_YX \equiv [Y,X]=\left(\frac{k}{2}-1\right)X+P\dd Z\,,\qquad\mbox{at}\qquad Z=0\,.
\end{equation}
As a consequence, the following bivector
\begin{equation}\label{p-p}
P'=X\wedge Y
\end{equation}
is invariant with respect to the flow of the Hamiltonian vector field $X=P\dd H$
\[
\mathcal L_X P'=[X,X]\wedge Y+X\wedge [X,Y]=0
\]
and satisfies the Jacobi identity 
\[
[\![P',P']\!]=-2X\wedge [Y,X]\wedge Y=0\,,
\]
which is a direct consequence of equation (\ref{y-eq}), valid only when the Lagrange identity (\ref{eq-lag}) holds.

Thus we have proved the following assertion.
\begin{prop}
If the Hamiltonian function $H=T+V$ (\ref{ham-1}) is an invariant of the weighted
dilation generator $Y$ (\ref{y-1}), that is,
\[Y(H)=k H\,,\] then the tensor field $P'$ (\ref{p-p}) is a Poisson bivector invariant with respect to the phase flow generated by the Hamiltonian vector field $X$ (\ref{x-vec}).
\end{prop}
In the variables $q$ and $p$ this Poisson bivector has the following form
\begin{align}
P'=&\sum_{1\leq i<j\leq n}\left(\frac{p_iq_j}{m_i}-\frac{p_jq_i}{m_j}\right)\frac{\partial}{\partial q_i}\wedge\frac{\partial}{\partial q_j}+ \sum_{i=1}^n\sum_{j=1}^n
\left(\frac{k}{2m_i}p_ip_j +q_i\frac{\partial V}{\partial q_j}\right)\frac{\partial}{\partial q_i}\wedge\frac{\partial}{\partial p_j}
\nonumber\\
+&\sum_{1\leq i<j\leq n} \frac{k}{2}
\left(
p_i\frac{\partial V}{\partial q_j}-p_j\frac{\partial V}{\partial q_i}
\right)\frac{\partial}{\partial p_i}\wedge\frac{\partial}{\partial p_j}\,.
\label{pp-ex}
\end{align}
For $k\neq -2$ and $n>1$, the invariant Poisson bivector $P'$ (\ref{p-p}) is not compatible with the Poisson bivector ${P}$ (\ref{p-can}), that is,
\[[\![P,P']\!]\neq 0\,.\]
Indeed, since
\[
\mathcal L_Y\,P=-\left(1+\frac{k}{2}\right)P\,,
\]
we have
\[
[\![P,P']\!]=(\mathcal L_X P)\wedge Y-(\mathcal L_Y P)\wedge X=
\left(1+\frac{k}{2}\right)X\wedge P\,.
\]
Recall that Poisson bivectors $P$ and $P'$ are compatible if any of their linear combinations $(a P+b P')$ with constant coefficients $a$ and $b$ (or, equivalently, simply their sum $P+P'$) is a Poisson bivector, that is, the Jacobi identity holds
\[
[\![a P+b P',a P+b P']\!]=2ab[\![P,P']\!]=0\,,\qquad \forall a,b\in \mathbb C.
\]
In the case of the Jacobi potential field, at $k=-2$ the compatibility condition $[\![P,{P}']\!]=0$ holds, and in this case there exists one more scalar invariant
\[
F=\frac{1}{2}\sum_{i,j=1}^n (q_ip_j-q_jp_i)^2 +2 V\sum_{i=1}^nq_i^2\,,\qquad \mathcal L_XF=0\,,
\]
see the definition of the Jacobi field and the discussion of the Lagrange identity in \cite{koz25}.

The Poisson bivector $P'$ (\ref{p-p}), $\rank P'=2$, is degenerate for $n\geq2$  and it satisfies the relation
\[
P'\dd H=kHX\,.
\]
Let us introduce non-degenerate invariant Poisson bivector
\[
\hat{P}=A(H)P+B(H) P'
\]
depending on two smooth functions $A(H)$ and $B(H)$ so that Jacobi identity holds 
\[
[\![\hat{P},\hat{P}]\!]=[\![AP+BP',\;AP+BP']\!]=0\,.\]
Applying the graded Leibnitz rule we obtain
\[
[\![\hat{P},\hat{P}]\!]=A^{2}\,[\![P,P]\!]+B^{2}\,[\![P',P']\!]+2AB\,[\![P,P']\!]
+2\,P\wedge \left(\hat{P}\dd A\right)+2\,P'\wedge \left(\hat{P}\dd{B}\right)\,.
\]
In our case 
\[
[\![P,P]\!]=[\![P',P']\!]=0\qquad\mbox{and}\qquad [\![P,P']\!]=
\left(1+\frac{k}{2}\right)X\wedge P\,,
\]
and, therefore, functions $A(H)$ and $B(H)$ have to satisfy equation
\begin{equation}\label{AB-poi}
2AB\left(1+\frac{k}{2}\right)X\wedge P
+2\,P\wedge \left(\hat{P}\dd A\right)  +2\,P'\wedge \left(\hat{P}\dd{B}\right)=0\,,
\end{equation}
where
\[
\hat{P}\dd A=\frac{\dd A}{\dd H}\, \hat{P}\dd H=\frac{\dd A}{\dd H}\, (A+kBH)X\qquad\mbox{and}\qquad
\hat{P}\dd B=\frac{\dd B}{\dd H}\, \hat{P}\dd H=\frac{\dd B}{\dd H}\, (A+kBH)X\,.
\]
Substituting $P'$ (\ref{p-p})  into the third term in (\ref{AB-poi}) one gets
\[
P'\wedge \left(\hat{P}\dd{B}\right)=\frac{\dd B}{\dd H}\, (A+kBH) X\wedge Y\wedge X=0\,,
\] 
that reduces  (\ref{AB-poi}) to the ordinary differential equation
\[
AB\left(1+\frac{k}{2}\right)-(A+kBH)\frac{\dd A}{\dd H}=0\,.
\]
One of the partial solution is 
\[
A=\left(1 - \frac{k}{2}\right)H\qquad\mbox{and}\qquad B=1\,.
\] 
Since $\rank P=2n$ the Poisson bivector $\hat{P}=AP+BP'$ has the same rank at $A(H) \neq 0$.

\begin{prop}
If $V(q)$ is a homogeneous function of homogeneity degree $k\neq \pm2$, then the tensor field of valence $(2,0)$
\begin{equation*}
\hat{P}=\left(1 - \frac{k}{2}\right)H P + P'\,,\qquad \rank\hat{P}=2n\,,
\end{equation*}
is a non-degenerate Poisson bivector on the open subset of the phase space defined by the inequality $H\neq0$ and $k\neq 2$. This allows one to define on this open subset a tensor field of valence $(0,2)$
\[
\hat{\omega}=\hat{P}^{-1}\,,
\]
which is a symplectic form invariant with respect to the phase flow generated by the Hamiltonian vector field $X$ (\ref{x-vec}).
\end{prop}

According to the Darboux theorem, at least locally there exists a coordinate transformation $\phi:(q,p)\to(\hat{q},\hat{p})$ such that
\[
\hat{P}=\sum_{i=1}^n\frac{\partial}{\partial \hat{q}_i}\wedge\frac{\partial}{\partial \hat{p}_i}\qquad \mbox{and}\qquad
\hat{\omega}=\sum_{i=1}^n d\hat{p}_i\wedge d\hat{q_i}\,.
\]
This transformation can be regarded as an analogue of the B\"{a}cklund transformations for integrable Hamiltonian systems with homogeneous potentials \cite{ts15}.

For all three invariant Poisson bivectors $P$, $P'$, and $\hat{P}$, the flows generated by $X$ are proportional to each other 
\[PdH=X\,,\qquad P'dH=kHX\qquad \mbox{and} \qquad  \hat{P}dH=\left(1+\frac{k}{2}\right)H\,X\,,\]
that makes the flow Hamiltonian with Hamilton functions $H$ and $\ln|H|$ respectively.  

\subsection{Absolute integral invariants}
According to Cartan's general homotopy formula, the Lie derivative $\mathcal L_X$ acting on a differential $k$-form $\varphi$ is equal to
\begin{equation}\label{hom-eq}
\mathcal L_X\varphi=\iota_X \dd\varphi+\dd\iota_X\varphi\,.
\end{equation}
Here $\iota_X\varphi$ is the interior product of the vector $X$ and the form $\varphi$, that is, the $(k - 1)$-form obtained from the form $\varphi$ by substituting the vector field $X$
in place of the first argument.

Following to Poincar\'{e} and Cartan \cite{poi,car}, a differential form $\varphi$ is an absolute integral invariant if both terms in (\ref{hom-eq}) simultaneously vanish
\[
\iota_X \dd\varphi=0\qquad\mbox{and}\qquad \dd \iota_X\varphi=0\,.
\]

In Hamiltonian mechanics, a canonical transformation $(q,p)\to (Q,P)$ can be generated by functions of mixed old/new variables. One of the standard generating function is $F(q,P)$ so that   
\[
\dd F=\sum_{i=1}^n p_i\dd q_i+\sum_{i=1}^n Q_i\dd P_i\,.
\] 
The  weighted dilation transformation (\ref{y-1}) is defined by differential 1-form 
\[\alpha_w =\iota_Y\omega\qquad\mbox{so that}\qquad \dd\alpha_w=\bigl(\tfrac{k}{2}+1\bigr)\omega\,.
\]
In local coordinates it has the following form 
\begin{equation*}
\alpha_w=\frac{k}{2}\alpha+\beta\,,\qquad \alpha=\sum_{i=1}^n p_i\dd q_i\,,\quad \beta=-\sum_{i=1}^n q_i\dd p_i\,.
\end{equation*}
The Lagrange identity (\ref{euler}) appears in this context as the weight identity
  \[\alpha_w(X)=kH\,,\qquad\mbox{when}\qquad Z=0\,.\] Together with 
\[\iota_X \dd\alpha_w=\left(\frac{k}{2}+1\right)\iota_X\omega=-\left(\frac{k}{2}+1\right)\dd H
\]  
Cartan's homotopy formula gives
\[
\mathcal L_X\alpha_w=\iota_X \dd\alpha_w+\dd \iota_X \alpha_w=\left(\frac{k}{2}-1\right)\dd H\,.
\]
The flow of $X$ does not preserve $\alpha_w$ in general, but drags it along $\dd H$.
It means that $\left.\alpha_w\right|_{H=h}$  is an absolute integral invariant on every energy level since
$\dd H=0$ on a fixed energy level.

The flow preserves the rank-two 2-form
\begin{equation}\label{wp}
\omega'=\dd H\wedge\alpha_w\,,
\end{equation}
since
\[
\mathcal L_X\omega'=\left(\mathcal L_X \dd H\right)\wedge\alpha_w+
\dd H\wedge\left(\mathcal L_X\alpha_w\right)=0\,.\]
This form is not closed for $k\neq -1$ and $n>1$ in view of
\[
\dd\omega'=-\dd H\wedge \dd \alpha_w=-\left(\frac{k}{2}+1\right)\dd H\wedge\omega\neq 0\,.
\]
\begin{prop}
The linear combination of the invariant 2-forms $\omega$ and $\omega'$
\begin{equation}\label{w-ab}
\omega_{ab}=a\omega+b\omega'
\end{equation}
with smooth coefficients $a=a(H)$ and $b=b(H)$ depending on $H$ is an absolute integral invariant. 
At 
\begin{equation}\label{w-cl}
a'(H)=(1+k/2)b(H)
\end{equation}
the invariant 2-form $\omega_{ab}$ (\ref{w-ab}) is closed.
\end{prop}
The proof consists in a direct verification of the Cartan conditions
\[
\iota_X\,\dd\omega_{ab}=\left(a'+\left(1+\frac{k}{2}\right)b\right)\iota_X \dd H\wedge \omega=0\,,
\]
where $a'=\dd a(H)/\dd H$, and
\[
\dd\,\iota_X\omega_{ab}=-\dd\Bigl(\left(a+kHb\right) \dd H\Bigr)=0\,.
\]
If (\ref{w-cl}) holds, then
\[
\dd\omega_{ab}=d(a\omega+b\dd H\wedge\alpha_w)=\bigl(a'-(1+k/2)b\bigr)\dd H\wedge\omega=0\,.
\]

The corresponding invariant bivectors
\[
P_{ab}=\omega_{ab}^{-1}=\frac{P}{a}-\frac{bP'}{a(a+kbH)}\,,
\]
where $P'=X\wedge Y$ (\ref{p-p}) are defined wherever the denominators of the fractions do not vanish. We can now relate functions $a(H)$, $b(H)$ with the functions $A(H)$, $B(H)$.

\subsection{Example: non-integrable H\'{e}non-Heiles system}
Let us consider the following polynomial Hamiltonian 
\[
H=\frac{p_1^2+p_2^2}{2}+q_1^M(c_1 q_1^N + c_2 q_2^N)\,,\qquad c_1,c_2\in\mathbb R\,.
\]
depending on arbitrary parameters $c_1$ and $c_2$ and a pair of integer $M$ and $N$.

In this case invariant Poisson bivector $P'$ is given by (\ref{pp-ex}) whereas
 the corresponding invariant symplectic form  $\omega'$ (\ref{wp}) in local coordinates is equal to 
\begin{align*}
{\omega'}=&\left(
\tfrac{N(M + N)c_2q_1^Mq_2^{N - 1}p_1}{2} - \tfrac{(M+N)\bigl((M+N)c_1q_1^N + Mc_2q_2^N\bigr)q_1^{M - 1} p_2}{2}
\right)\dd q_1\wedge \dd q_2\\
+&\left(
\tfrac{M + N}{2}p_1^2 + q_1^M\bigl((M+N)c_1q_1^N + Mc_2q_2^N\bigr)
\right)\dd q_1\wedge \dd p_1\\
+&\left(\tfrac{M + N}{2}p_1p_2 + q_2q_1^{M - 1}\bigl((M+N)c_1q_1^N + Mc_2q_2^N\bigr)\right)\dd q_1\wedge \dd p_2\\
+&\left(
\tfrac{M + N}{2}p_1p_2 + Nc_2q_1^{M+1}q_2^{N - 1}\right)\dd q_2\wedge \dd p_1\\
+&\left(\tfrac{M + N}{2}p_2^2 + Nc_2q_1^Mq_2^N\right)\dd q_2\wedge \dd p_2
+\left(p_1q_2 - p_2q_1\right) \dd p_1\wedge \dd p_2\,.
\end{align*}
At $M=1$ and $N=2$ we have a famous H\'{e}non-Heiles which is integrable only for special values of $c_1$ and $c_2$.

It will be interesting to construct geometric or symplectic integrator for the non-integrable H\'{e}non-Heiles system preserving both symplectic forms $\omega$ and $\omega'$ simultaneously. 

\section{Non-standard family of weight-homogeneous vector fields}
\setcounter{equation}{0}
Consider the Hamiltonian function
\[
H=\frac{1}{2m_1} p_1^2+\cdots+\frac{1}{2m_n} p_n^2+U(q)\,,
\]
in which the potential $U(q)$ satisfies the relation
\begin{equation}\label{euler2}
\bar{Z}=0\,,\qquad\mbox{where}\qquad \bar{Z}=\sum_{i=1}^n a_i\frac{\partial}{\partial q_i}\,U(q_1,\ldots,q_n)-kU(q_1,\ldots,q_n)\,.
\end{equation}
In this case, the analogue of the Lagrange identity (\ref{eq-lag}) is the expression
\begin{equation*}
\frac{d^2}{dt^2} \sum_{i=1}^n a_iq_i=-kU.
\end{equation*}
A particular solution of the Euler equation (\ref{euler}),
\[
V(q_1,\ldots, q_n)=q_1^kf\left(\frac{q_2}{q_1},\ldots,\frac{q_n}{q_1}\right)\,,
\]
where $f$ is an arbitrary smooth function of the projective coordinates, can be compared with a similar particular solution of equation (\ref{euler2}), which has the form
\[
U(q_1,\ldots, q_n)=e^{\frac{k}{a_1}q_1}g\left(
q_2-\frac{a_2}{a_1}\,q_1,\ldots, q_n-\frac{a_n}{a_1}\,q_1\right)\,,
\]
where $g$ is an arbitrary smooth function. A particular case of Hamiltonian systems with potential $U(q)$ is given by some of the Toda lattices \cite{ts25c}.

Let us define the vector field
\begin{equation}\label{y-2}
\bar{Y}=\sum_{i=1}^n a_i\frac{\partial}{\partial q_i}+\frac{k}{2}
\sum_{i=1}^n p_i\frac{\partial}{\partial p_i}\,,
\end{equation}
for which the sum $H=T+U$ is an invariant provided that the condition $\bar{Z}=0$ holds
\[
\bar{Y}(T+U)=k(T+U)+\bar{Z}=k(T+U)\,.
\]
The corresponding vector field
 \[
\bar{X}=P\dd H=\left(\sum_{i=1}^n \frac{p_i}{m_i} \frac{\partial}{\partial q_i}\right)-
\left(\sum_{j=1}^n \frac{\partial U}{\partial q_j}\dfrac{\partial}{\partial p_j}\right)
\]
is also an invariant of the vector field $\bar{Y}$ (\ref{y-2})
\[
[\bar{Y},\bar{X}]=\frac{k}{2}\bar{X}+Pd\bar{Z}\qquad\mbox{provided that}\qquad \bar{Z}=0\,.
\]
This immediately implies the validity of the following assertion.
\begin{prop}
The rank-two bivector
\[\bar{P}'=\bar{X}\wedge\bar{Y}\]
is a Poisson bivector invariant with respect to the phase flow of the vector field $\bar{X}$ if the function $U(q)$ satisfies equation (\ref{euler2}), that is, if
\[\bar{Y}(H)=kH\,.\]
\end{prop}
The proof, as before, consists in substituting the Lie bracket between the vector fields
$\bar{X}$ and $\bar{Y}$ into the equations
\[\mathcal L_{\bar{X}}\,\bar{P}' =[\bar{X},\bar{X}]\wedge \bar{Y}+\bar{X}\wedge [\bar{X},\bar{Y}]=0\]
and
\[
[\![\bar{P}',\bar{P}']\!]=-2\bar{X}\wedge [\bar{Y},\bar{X}]\wedge \bar{Y}=0\,.
\]
In the variables $q$ and $p$ this Poisson bivector has the following form
\begin{align}
\bar{P}'=&\sum_{1\leq i<j\leq n}\left(\frac{a_jp_i}{m_i}-\frac{a_ip_j}{m_j}\right)\frac{\partial}{\partial q_i}\wedge\frac{\partial}{\partial q_j}+
\sum_{i=1}^n\sum_{j=1}^n
\left(\frac{k}{2m_i}p_ip_j +a_i\frac{\partial U}{\partial q_j}\right)\frac{\partial}{\partial q_i}\wedge\frac{\partial}{\partial p_j}\nonumber\\
+&\sum_{1\leq i<j\leq n} \frac{k}{2}
\left(
p_i\frac{\partial U}{\partial q_j}-p_j\frac{\partial U}{\partial q_i}
\right)\frac{\partial}{\partial p_i}\wedge\frac{\partial}{\partial p_j}\,.
\label{p-p2}
\end{align}
The Poisson bivector $\bar{P}'$ (\ref{p-p2}), $\rank \bar{P}'=2$, is degenerate for $n\geq2$ and satisfies the relation
\[
\bar{P}'\dd H=kH\bar{X}\,.
\]
As above we can introduce non-degenerate invariant Poisson bivector
\[
\bar{P}=A(H)P+B(H) \bar{P}'
\]
depending on two smooth functions $A(H)$ and $B(H)$ so that Jacobi identity has the form  
\[
[\![\bar{P},\bar{P}]\!]=2AB\,[\![P,\bar{P}']\!]
+2\,P\wedge \left(\bar{P}\dd A\right)=0\,.
\]
At $B=1$ this equation has the following solution
\[
A(H)=- \frac{k}{2} H\,,
\]
and $\rank \bar{P}=2n$ at $H\neq 0$.
\begin{prop}
If the potential $U(q)$ satisfies relation (\ref{euler2}) with $k\neq 0$, then on the open subset defined by the inequality $H\neq0$, the tensor field of valence $(2,0)$
\begin{equation*}
\bar{P}= - \frac{k}{2} H P + \bar{P}'\,,\qquad \rank\bar{P}=2n\,,
\end{equation*}
is a non-degenerate invariant Poisson bivector, and the tensor field of valence $(0,2)$
\[
\bar{\omega}=\bar{P}^{-1}
\]
is a symplectic form invariant with respect to the flow generated by the Hamiltonian vector field $\bar{X}$.
\end{prop}

Using symplectic form $\bar{\alpha}_w =\iota_{\bar{Y}}\omega$ we can obtain  invariant symplectic forms
\[
\bar{\omega}'=\dd H\wedge\bar{\alpha}_w\,,\qquad \bar{\omega}_{ab}=aw+ b\bar{\omega}'
\]
and other absolute invariants as in the previous case.

\subsection{Example: 2D Hamiltonian system with exponential potential }
If the Hamiltonian vector field $X$ (\ref{x-vec}) is defined by potential 
\[
U(q_1,q_2)=\exp\left(\frac{-4q_1}{\beta}\right)F\left(\frac{\alpha q_1 + \beta q_2}{\beta}\right)
\] 
then the corresponding flow preserves the canonical Poisson bivector $P$ (\ref{p-can}) 
and  bivector $\bar{P}'$ (\ref{p-p2})
\[
\bar{P}'=\left(
                  \begin{smallmatrix}
                    0 & \alpha p_1 + \beta p_2 & p_1^2 - p_2^2 - \alpha\partial_2 U + 2U & 2p_1p_2 - \beta \partial_2 U \\ 
                    0 & 0 & 2p_1p_2 + \alpha \partial_1 U & -p_1^2 + p_2^2 + \alpha\partial_2 U - 2U \\ 
                    * & * & 0 &2p_1\partial_2 U p_1 - 2p_2\partial_1U   \\ 
                   *  & * & * & 0 
                  \end{smallmatrix}
                \right)\,.
\] 
Here $\alpha,\beta\in\mathbb R$  are arbitrary parameters and $F\left((\alpha q_1 + \beta q_2)/\beta\right)$ is an arbitrary function.

The best-known examples of such type potentials are open Toda lattice ($\alpha=0$) and periodic Toda lattice ($\alpha=1$) associated with the root system $G_2$
\[
H=\frac{p_1^2 + p_2^2}{2} + \exp(q_1/\sqrt{3}) + \exp(-\sqrt{3}/2q_1 + q_2/2) + \alpha\exp(-q_2)\,.
\]

\section{Conclusion}
The modern version \cite{lee-08} of the Clebsch, Mayer, Lie and Frobenius theorem has the form:\\
\textit{"A smooth distribution $D\subset TM$ is integrable if and only if it is involutive.
Moreover, in that case every point has a neighbourhood $U$ with coordinates
$(x_1,\dots,x_k,y_1,\dots,y_{n-k})$ such that
\[
D = \operatorname{span}\left\{\frac{\partial}{\partial x_1},\dots,
\frac{\partial}{\partial x_k}\right\} \quad\text{on } U,
\]
and the integral submanifolds in $U$ are the plaques $y_1=c_1,\dots,y_{n-k}=c_{n-k}$.
In particular, an integrable distribution partitions $M$ into maximal connected
integral submanifolds."
}

We apply this theorem to construct global tensor invariants for homogeneous and scale-invariant vector fields. Furthermore, we demonstrate that analogous additional tensor invariants exist for certain weight-homogeneous decompositions.

 The following conceptual questions remain open: tensor invariants on the singular locus $\{H=0\}$, tensor invariants at degenerate cases $k=\pm 2$, tensor invariants at the fixed point of $X$ where the fields are dependent, etc. Note that the cases where $k = \pm 2$ are special cases within the framework of the application of Galois differential group theory to Hamiltonian systems with homogeneous potentials \cite{goriely}.

\subsection*{Acknowledgments} {The article was prepared within the framework of the project “International academic cooperation” at HSE University.}
The author thanks A.~Maciejewski for comments on the original version of this paper, which made it possible to substantially improve it.

\

\begin{thebibliography}{99}

\bibitem{alb}
A. Albouy, H. Cabral, A. Santos, Some problems on the classical $n$-body problem,
\newblock{\em Celest. Mech. Dyn. Astr.}, \textbf{113} (2012), 369-375.

\bibitem{bog96}  
O.\,I. Bogoyavlenskij, 
\newblock{Theory of tensor invariants of integrable Hamiltonian systems. I. Incompatible Poisson structures},
\newblock{\em Commun. Math. Phys.}, \textbf{180}, (1996), 529-586.

\bibitem{bog97} 
O.\,I. Bogoyavlenskij,
\newblock{Theory of tensor invariants of integrable Hamiltonian systems. II. Theorem on
symmetries and its applications}, 
\newblock{\em   Commun. Math. Phys.}, \textbf{184} (1991), 301-365.

\bibitem{bri}
F. Brioschi, Sur un syst\`{e}me d’\'{e}quations diff\'{e}rentielles, 
\newblock{\em C. R. Acad. Sci. Paris}, \textbf{92} (1881), 1389-1393.

\bibitem{car}
 E. Cartan, \newblock{\em Le\c{c}ons sur les invariants int\'{e}graux}, Hermann, Paris, 1922, 210 pp.

\bibitem{cl-66}
A. Clebsch,
{\"Uber die simultane Integration linearer partieller Differentialgleichungen},
\newblock{\emph J.~Reine Angew.~Math.}, \textbf{65} (1866), 257-268.

\bibitem{deah-40}
H. Deahna,  {\"Uber  die Bedingungen der Integrabilit\"{a}t linearer Differentialgleichungen
erster Ordnung zwischen einer beliebigen Anzahl ver\"{a}anderlicher Gr\"{o}ossen},
\newblock{\em Journal f\"{u}r die reine und angewandte Mathematik},
\textbf{20} (1840), 340-349.

\bibitem{eul-int}
L. Euler, \newblock{\em Institutiones Calculi Integralis volumen primum}, Acta Petropolitana, 1761.

\bibitem{forth-06}
A.R. Forsyth, \newblock{\em Theory of Differential Equations}, vol 1-6, Cambridge University Press (1890-1906). 

\bibitem{frob-77}
G. Frobenius, \"{U}ber das Pfaffsche Problem, \newblock{\em Journal f\"{u}r die reine und angewandte Mathematik}, \textbf{82} (1877), 230-315.



\bibitem{goriely}
A.~Goriely,
\emph{Integrability and Nonintegrability of Dynamical Systems},
Advanced Series in Nonlinear Dynamics \textbf{19}, World Scientific, 2001.

\bibitem{jac-27}
C.G.J. Jacobi,  \"{U}ber die Integration der partiellen Differentialgleichungen erster Ordnung,
\newblock{\em  Journal f\"{u}r die reine und angewandte Mathematik}, \textbf{2} (1827), 317-329.

\bibitem{jac-27a}
C.G.J. Jacobi,  \"{U}ber die PfaffscheMethode, eine gew\"{o}hnliche lineare Differentialgleichung
zwischen $2n$ Variabeln durch ein System von $n$ Gleichungen zu integriren,
\newblock{\em  Journal f\"{u}r die reine und angewandte Mathematik},
\textbf{2} (1827), 347-357.


\bibitem{jac}
Jacobi C.G.J., Theoria novi multiplicatoris systemati {\ae}quationum differentialium vulgarium
applicandi,  \newblock{\em Journal f\"{u}r die reine und angewandte Mathematik}, \textbf{27} (1844), 199-268, \textbf{29} (1845), 213-279 and 333-376.

\bibitem{kl-74}
G. H. Katzin, J. Levine,
\newblock{Dynamical symmetries and constants of the motion for classical particle systems}
\newblock{\em J. Math. Phys.}, \textbf{15} (1974), 1460-1470.


\bibitem{koz-24}
V.V. Kozlov, \newblock{Solvable algebras and integrable systems},
\newblock{\em  Regul. Chaot. Dyn.}, \textbf{29}:5 (2024), 717-727.

\bibitem{koz25}
V.V. Kozlov,  The Lagrange Identity and Dynamics
in a Potential Jacobi Field, {\em Regul. Chaot. Dyn.}, \textbf{30}:6 (2025), 887–907. 

\bibitem{lee-08}
J.M. Lee, Distributions and Foliations. In: \newblock{\em  Introduction to Smooth Manifolds}, Graduate Texts in Mathematics, \textbf{ 218}, Springer, New York, (2013).

\bibitem{lie-1}
S. Lie, Verallgemeinerung und neue Verwertung der Jacobischen Multiplikatortheorie, 
\newblock{\em Forhandlinger Christiania 1874}, (1875), 255-274. 

\bibitem{lie-77}
S. Lie, Theorie des Pfaffschen Problems I, \newblock{\em  Archiv for Mathematik}, \textbf{2} (1877), 338-379. 

\bibitem{lie-book}
S. Lie, \newblock{\em Theorie der Transformationsgruppen. Abhandlung I}, unter Mitwirkung von Dr. Friedrich Engel, Leipzig, 1888.

\bibitem{llibre}
J. Llibre, X.Zhang,  On the integrability of the Hamiltonian systems with homogeneous polynomial potentials,
\newblock{ \em Applied Mathematics and Nonlinear Sciences}, \textbf{3}:2 (2018), 527-536.


\bibitem{adam}
A.J. Maciejewski, M. Przybylska,  Partial integrability of Hamiltonian systems with homogeneous potential,
\newblock{\em  Regul. Chaotic Dyn.}, \textbf{15}:4 (2010), 551-563.

\bibitem{may-72}
A. Mayer. \"{U}ber unbeschr\"{a}nkt integrable Systeme von linearen totalen Differentialgleichungen
und die simultane Integration linearer partieller Differentialgleichungen,
\newblock{\em  Math. Ann.}, \textbf{1872}, 448-470.

\bibitem{mm-21}
M. Markoutsakis, \newblock{\em Geometry, Symmetries, and Classical Physics: A Mosaic}, CRC Press, (2021).



\bibitem{no-18}
E. Noether, \newblock{\em Invariante Variationsprobleme}, Nachrichten der Königlichen Gesellschaft der Wissenschaften zu G\"{o}ttingen, Mathematisch-Physikalische Klasse, (1918), 235-257.




\bibitem{pfaff-15}
 J.~F. Pfaff,
\newblock{Methodus generalis, aequationes differentiarum partialium, nec non aequationes
differentiales vulgares, utrasque primi ordinis, inter quotcumque variabiles, complete
integrandi},
\newblock{\em Abhandlungen der K\"oniglichen Akademie der Wissenschaften in Berlin}, (1814-1815),
76-136.

\bibitem{poi}
H. Poincar\'{e},
\newblock{\em Les M\'{e}thodes Nouvelles de la M\'{e}canique C\'{e}leste}, Tom III, Dover Pub. Inc., N.Y., 1892.

\bibitem{ts15}
A.\,V. Tsiganov,
\newblock{On auto and hetero B{\"a}cklund transformations for the H{\'e}non-Heiles systems},
\newblock{\em Phys. Lett. A}, \textbf{379} (2015),  2903-2907.


\bibitem{ts25} 
A.\,V. Tsiganov, On tensor invariants for integrable cases of Euler, Lagrange and Kovalevskaya rigid body motion,
\newblock{\em  Izv. Math.}, \textbf{89}:2 (2025), 372-398.

  
\bibitem{ts25a}
A.\,V. Tsiganov, { On tensor invariants of the Clebsch system}, 
\newblock{\em  Regul. Chaotic Dyn.}, \textbf{30}:4 (2025) 742-764.
 
 
  \bibitem{ts25b}
A.\,V. Tsiganov, 
{ On novel Hamiltonian description of the nonholonomic Suslov problem},
\newblock{\em  Regul. Chaotic Dyn.}, \textbf{31} (2026), 263–272.  

 
\bibitem{ts25c}
A.\,V. Tsiganov, {Experimental study of tensor invariants of Hamiltonian systems},
\newblock{\em  Theor. Appl. Mech.}, \textbf{52}:2 (2025), 297-316.

\bibitem{ts25ar}
Tsiganov A.V., Homogeneous potentials, Lagrange's identity and Poisson geometry, Preprint: arXiv:2511.19903, (2025).
(2026).

\bibitem{ts26a}
A.\,V. Tsiganov, On formal Hamiltonian description of homogeneous vector fields, 
{\em Russ. Math. Surv.}, \textbf{81}:2 (2026), 379-381.

\bibitem{ts26d}
Tsiganov A.V., On the Darboux-Halphen system: Jacobi vs Lie, 
\newblock{\em J. Geom. Phys.}, 228 (2026), 105914.

\bibitem{ts26rcd}
Tsiganov A.V., On the Lie noncommutative integrability,
\newblock{\em Regul. Chaotic Dyn.}, \textbf{31}:6 (2026), 928-943.



\end{thebibliography}
\end{document}